\documentclass[12pt]{article}
\usepackage{amsfonts,amssymb,amsmath}
\usepackage[dvips]{epsfig}
\begin{document}
\title{\bf Protection of entanglement for a two-qutrit V-type open system on the basis of system-reservoir bound states}
\author{B. Ahansaz $^{a}$, N. Behzadi $^{b}$ \thanks{E-mail:bahramahansaz@gmail.com}
and E. Faizi $^{a}$,
\\ $^a${\small Physics Department, Azarbaijan Shahid Madani University, Tabriz, Iran,}
\\ $^b${\small Research Institute for Fundamental Sciences, University of Tabriz, Tabriz, Iran}} \maketitle

\begin{abstract}
\noindent
In this work, a mechanism for long time protection of entanglement of a two-qutrit V-type atomic system each of which interacts independently with a dissipative reservoir is investigated. It is shown that the protection process is related exclusively on the formation of bound states in system-reservoir energy spectrum. It is demonstrated that the degree of boundedness for the system-reservoir bound states is well-controlled by entering other V-type qutrits into the reservoirs, and also through the manipulation of the spontaneously generated interference related to decay channels of the V-type atoms.
\\
\\
{\bf PACS Nos:}
\end{abstract}

\section{Introduction}
Quantum entanglement is a fundamental concept in quantum mechanics, which has broad applications in various quantum
information processing tasks such as quantum key distribution, quantum teleportation, quantum dense coding and quantum computation \cite{Horodecki}.
Since every natural physical quantum object is in contact with its surroundings environment, so a consequence of this resulting inevitable
interaction is decoherence which leads to the destruction of entanglement.
Therefore, researchers used several strategies to preserve entanglement for a long time limit to allow the performances of quantum computation and quantum communication.
In this regard, it was shown that the evolution of a quantum system can be slowed down or even frozen by frequent repeated projective measurements
during a defined period of time which is known as the quantum Zeno effect $\cite{Maniscalco, Mundarain, Rossi, Hou}$.
Also, the other way for protecting the entanglement was performed by modulating the detuning between the qubits and the central frequency
of an imperfect cavity $\cite{Bellomo, Xiao}$.
One of the interesting strategies for the protection of the long time limit of quantum entanglement was observed in $\cite{An2}$, where it was shown that
the bipartite entanglement can be sustained by the addition of a third qubit. The extension of this scheme, by considering more additional qubits in the dissipative reservoir, was introduced in $\cite{An3}$.

On the other hand, entanglement protection between two multilevel atoms using addition of other multilevel atoms, as used for protection of two-qubit entanglement, has been not performed yet.
Multilevel atomic systems have important applications in quantum key distribution \cite{Bourennane} and
in the increment of violation of the local realism \cite{Kaszlikowski}.
Moreover, it was demonstrated that the quantum interference between possible transitions occurred in multilevel atomic systems can lead to many novel and interesting effects such as
absorption reduction and cancellation \cite{Harris,Scully} and spontaneous emission reduction and cancellation \cite{Scully}.

Based on the previous results, it was found that the formation of the system-environment bound states, as special eigenstates with eigenvalues residing in the band gap of the energy spectrum, significantly changes the dynamics of the few-level atoms. They indeed play an important role in exhibiting a fractional steady-state atomic population known as population trapping $\cite{Lambropoulos}$ as well as entanglement protection $\cite{Tong}$.

In this paper, we are interested to investigate the protection of entanglement between two distant V-type three-level atoms considered as qutrits, each of which is contained independently in a dissipative reservoirs, in the presence of other $N-1$ similar V-type atoms. We demonstrate that the initial two-qutrit entanglement can be well-protected exclusively by improving the degree of boundedness of the system-reservoir bound states. On the other hand, we show that the degree of boundedness for the system-reservoir bound states can be well-controlled by two manner successfully. First, by entering the other similar V-type atoms into the reservoirs and second, through the manipulation of the spontaneously generated interference (SGI) of decay channels of the V-type atoms.

The work is organized as follows: In Sec. II, we obtained the dynamics of a single-qutrit V-type open system in the presence of other similar qutrits analytically.
Sec. III is devoted to deriving the condition for the formation of bound states in the energy spectrum of the system-reservoir or total system.
In Sec. IV, controlling the entanglement dynamics between two distant V-type qutrits for protection of it from dissipation is analyzed, by considering that each atom is contained independently in a dissipative reservoir.
And finally, the paper is ended with a brief conclusion in Sec. V.

\section{Dynamics of a single-qutrit V-type open system in the presence of other qutrits}
In this section, we study the dynamics of a single-qutrit V-type atom as our open system in the presence of other additional V-type systems where all of them are coupled to a common zero-temperature thermal reservoir, as depicted in Fig. 1(a).
Each atom has two excited states $|A\rangle$ and $|B\rangle$ which can spontaneously decay into ground state $|C\rangle$ with transition frequencies $\omega_{A}$
and $\omega_{B}$ respectively. It should be noted that the respective dipole moments of transitions may have interaction with each other.
The Hamiltonian for the system can be written as $H=H_{0}+H_{I}$, where the free Hamiltonian $H_{0}$ is given by ($\hbar=1$)
\begin{eqnarray}
H_{0}=\sum_{l=1}^{N} \sum_{m=A,B} \omega_{m} \sigma_{m}^{l+} \sigma_{m}^{l-} +\sum_{k}\omega_{k}b_{k}^{\dagger}b_{k},
\end{eqnarray}
and the interaction Hamiltonian $H_{I}$ is
\begin{eqnarray}
H_{I}=\sum_{l=1}^{N} \sum_{m=A,B} \sum_{k} \big(g_{mk} \sigma_{m}^{l+} b_{k}+g_{mk}^{\ast} \sigma_{m}^{l-} b_{k}^{\dagger} \big).
\end{eqnarray}
In Eqs. (1) and (2), ${\sigma_{m}^{l\pm}}(m=A, B)$ are the raising and lowering operators of the $m^{th}$ excited state for the $l^{th}$ atom. Also, $b_{k}$ ($b_{k}^{\dagger}$) is the annihilation (creation) operator of the $k^{th}$ field mode with frequency $\omega_{k}$
and the strength of coupling between the $m^{th}$ excited state and the $k^{th}$ field mode of all atoms is the same and is given by $g_{mk}$. The Schr\"{o}dinger equation in the interaction picture is as
\begin{eqnarray}
i\frac{d}{dt}|\psi(t)\rangle=H_{int}(t)|\psi(t)\rangle,
\end{eqnarray}
where $H_{int}(t)=e^{i H_{0}t} H_{I} e^{-i H_{0}t}$ is given by
\begin{eqnarray}
H_{int}=\sum_{l=1}^{N} \sum_{m=A,B} \sum_{k} \big(g_{mk} \sigma_{m}^{l+} b_{k}e^{i(\omega_{m}-\omega_{k})t}+g_{mk}^{\ast} \sigma_{m}^{l-} b_{k}^{\dagger}e^{-i(\omega_{m}-\omega_{k})t}\big).
\end{eqnarray}
It is supposed that the initial state of the total system is as
\begin{eqnarray}
  |\psi(0)\rangle=\zeta_{0}(0)|0\rangle_{S} \otimes |0\rangle_{E}+\sum_{l=1}^{N} \big( \zeta_{l}^{A}(0)|A_{l}\rangle+ \zeta_{l}^{B}(0)|B_{l}\rangle \big)_{S} \otimes |0\rangle_{E},
\end{eqnarray}
Since the number of excitations in total system is conserved then the time-evolved state $|\psi(t)\rangle$ can be written as
\begin{eqnarray}
  |\psi(t)\rangle=\zeta_{0}(t)|0\rangle_{S} \otimes |0\rangle_{E}+\sum_{l=1}^{N} \big( \zeta_{l}^{A}(t)|A_{l}\rangle+ \zeta_{l}^{B}(t)|B_{l}\rangle \big)_{S} \otimes |0\rangle_{E}
  +\sum_{k} \nu_{k}(t) |0\rangle_{S} |1_{k}\rangle_{E}.
\end{eqnarray}
Here, we note that $|0\rangle_{S}=|C\rangle^{\otimes N}$, which means that all of the atoms are in the ground state
and also, $|A_{l}\rangle_{S}$ ($|B_{l}\rangle_{S}$) represents that all of the atoms are in the ground state except the $l^{th}$ atom which is in the corresponding excited state A (B).
Moreover, we denote $|0\rangle_{E}$ being the vacuum state of the reservoir and $|1_{k}\rangle_{E}$ is the state which has only one excitation in the $k^{th}$ mode. It is clear that $H_{int}(t) |0\rangle_{S} \otimes |0\rangle_{E}=0$ then $\zeta_{0}(t)=\zeta_{0}(0)=\zeta_{0}$. Substituting Eqs. (4) and (6) into Eq. (3), gives the following differential equations
$$
\dot\zeta_{l}^{A}(t)=-i \sum_{k} g_{Ak} e^{i(\omega_{A}-\omega_{k})t} \nu_{k}(t),
$$
\begin {equation}\label{}
\dot\zeta_{l}^{B}(t)=-i \sum_{k} g_{Bk} e^{i(\omega_{B}-\omega_{k})t} \nu_{k}(t),
\end {equation}
$$
\dot{\nu}_{k}(t)=-i \sum_{m=A,B} g_{mk}^{*} e^{-i(\omega_{m}-\omega_{k})t} \sum_{l=1}^{N} \zeta_{l}^{m}(t).
$$
By integrating from the last part of Eq. (7) and substituting it into the remainder ones, we obtain the following set of closed integro-differential equations
\begin{equation}
\frac{d\zeta_{l}^{m}(t)}{dt}=-\sum_{n=A,B} \int_{0}^{t} f_{mn}(t-t') \sum_{j=1}^{N} \zeta_{j}^{n}(t') dt', \quad m=A, B.
\end{equation}
The kernels in Eq. (8) is related to the spectral density $J(\omega)$ of the reservoir as
\begin{eqnarray}
f_{mn}(t-t')=\int_{0}^{t}d\omega J_{mn}(\omega)e^{i(\omega_{m}-\omega)t-i(\omega_{n}-\omega)t'}.
\end{eqnarray}
We take a Lorentzian spectral density of the reservoir as
\begin{eqnarray}
J_{mn}(\omega)=\frac{1}{2\pi}\frac{\gamma_{mn}\lambda^{2}}{(\omega_{0}-\omega)^{2}+\lambda^{2}},
\end{eqnarray}
where $\lambda$ is the spectral width of the coupling and $\omega_{0}$ is the central frequency of the structured reservoir.
Also, $\gamma_{mm}=\gamma_{m}$ are the relaxation rates of the two upper excited states
and $\gamma_{mn}= \sqrt{\gamma_{m} \gamma_{n}} \theta$ with $m\neq n$ and $\theta\leq1$, is responsible for the SGI
between the two decay channels $|A\rangle \rightarrow |C\rangle$ and $|B\rangle \rightarrow |C\rangle$ for each atom.
The parameter $\theta$ depends on the angle between two dipole moments of the mentioned transitions where $\theta=0$ means that the dipole moments of the transitions are perpendicular to each other, which is corresponding to the case that there is no SGI between two decay channels. On the other hand, $\theta=1$ indicates that the two dipole moments are parallel, which is corresponding to the strongest SGI between two decay channels.

Taking the Laplace transform from both sides of Eq. (8) gives the following set of equations
\begin{equation}
p \zeta_{l}^{m}(p)-\zeta_{l}^{m}(0)=-\sum_{n=A,B} \mathcal{L}\{f_{mn}(t)\} \sum_{j=1}^{N} \zeta_{j}^{n}(p), \quad m=A, B.
\end{equation}
where $\zeta_{l}^{m}(p)=\mathcal{L} \{\zeta_{l}^{m}(t)\}=\int_{0}^{\infty} \zeta_{l}^{m}(t) e^{-pt} dt$, is the Laplace transform of $\zeta_{l}^{m}(t)$.
Clearly using Eq. (11), we have the following equalities
\begin{equation}
p \zeta_{1}^{m}(p)-\zeta_{1}^{m}(0)=...=p \zeta_{l}^{m}(p)-\zeta_{l}^{m}(0)=...=p \zeta_{N}^{m}(p)-\zeta_{N}^{m}(0).
\end{equation}
So by considering these equalities, Eq. (11) takes the following form
\begin{eqnarray}
p \zeta_{l}^{m}(p)-\zeta_{l}^{m}(0)=-\sum_{n=A,B} \mathcal{L}\{f_{mn}(t)\} \bigg(N \zeta_{l}^{n}(p)+\frac{1}{p}\sum_{j\neq l}^{N} \big(\zeta_{j}^{n}(0)-\zeta_{l}^{n}(0)\big) \bigg), \quad m=A, B.
\end{eqnarray}

In this paper, we consider the case in which the two upper atomic states are degenerated and the atomic transitions are in resonant with the central frequency of the reservoir, i.e
$\omega_{A}=\omega_{B}=\omega_{0}$. Under this consideration, we can assume that
$\gamma_{A}=\gamma_{B}=\gamma_{0}$ and $\gamma_{AB}=\gamma_{BA}=\gamma_{0} \theta$, so the kernels in Eq. (9) takes a simple form as
\begin{equation}
\begin{array}{l}
\displaystyle f_{AA}(t-t')=f_{BB}(t-t')=f(t-t')=\int_{0}^{t}d\omega J(\omega)e^{i(\omega_{0}-\omega)(t-t')},\\\\
\displaystyle f_{AB}(t-t')=f_{BA}(t-t')=f'(t-t')=\int_{0}^{t}d\omega J'(\omega)e^{i(\omega_{0}-\omega)(t-t')},
\end{array}
\end{equation}
where $J'(\omega)=\theta J(\omega)$. To solve Eq. (13), we define the new coefficients $\zeta_{l}^{\pm}(p)=\zeta_{l}^{A}(p) \pm \zeta_{l}^{B}(p)$, and rewrite Eq. (13) as
\begin{eqnarray}
p \zeta_{l}^{\pm}(p)-\zeta_{l}^{\pm}(0)=-\big[\mathcal{L}\{f(t)\} \pm \mathcal{L}\{f'(t)\} \big] \bigg(N \zeta_{l}^{\pm}(p)+\frac{1}{p}\sum_{j\neq l}^{N} \big(\zeta_{j}^{\pm}(0)-\zeta_{l}^{\pm}(0)\big) \bigg),
\end{eqnarray}
and therefore it yields
\begin{eqnarray}
\zeta_{l}^{\pm}(t)=\mathcal{G}_{\pm}(t)\zeta_{l}^{\pm}(0)-\frac{1-\mathcal{G}_{\pm}(t)}{N} \sum_{j\neq l}^{N} \big(\zeta_{j}^{\pm}(0)-\zeta_{l}^{\pm}(0)\big).
\end{eqnarray}
where $\mathcal{G}_{\pm}(t)=e^{-\lambda t/2} \big(\mathrm{cosh}{(\frac{D^{\pm}t}{2})}+\frac{\lambda}{D^{\pm}} \mathrm{sinh}{(\frac{D^{\pm}t}{2})}\big)$ and $D^{\pm}=\sqrt{\lambda^{2}-2\gamma_{0} (1 \pm \theta) \lambda N}$.
Consequently the amplitudes $\zeta_{l}^{A}(t)=\big(\zeta_{l}^{+}(t)+\zeta_{l}^{-}(t)\big)/2$ and $\zeta_{l}^{B}(t)=\big(\zeta_{l}^{+}(t)-\zeta_{l}^{-}(t)\big)/2$ are obtained analytically.

\section{System-reservoir bound states}
To find the energy spectrum of the total system discussed in the previous section, we solve the following eigenvalue equation
\begin{eqnarray}
 \mathcal{H}(t) |\psi(t)\rangle=E |\psi(t)\rangle,
\end{eqnarray}
where $\mathcal{H}(t)=e^{i H_{0}t} H e^{-i H_{0}t}=H_{0}+H_{int}(t)$ is the Hamiltonian of the total system in the interaction picture and $|\psi(t)\rangle$ is given by Eq. (6). Since $\mathcal{H}(t)|0\rangle_{S} \otimes |0\rangle_{E}=0$, so the above eigenvalue equation imposes that $\zeta_{0}(t)=0$ and also
the following set of $2N+1$ equations can be obtained
$$
\omega_{k} \nu_{k}(t)+\sum_{l=1}^{N} \big(g_{Ak}^{*} \zeta_{l}^{A}(t)+g_{Bk}^{*} \zeta_{l}^{B}(t)\big) e^{-i(\omega_{0}-\omega_{k})t} =E \nu_{k}(t),
$$
\begin {equation}
\omega_{0} \zeta_{l}^{A}(t)+\sum_{k} g_{Ak} e^{i(\omega_{0}-\omega_{k})t} \nu_{k}(t)=E \zeta_{l}^{A}(t), \quad l=1, 2,...,N,
\end {equation}
$$
\omega_{0} \zeta_{l}^{B}(t)+\sum_{k} g_{Bk} e^{i(\omega_{0}-\omega_{k})t} \nu_{k}(t)=E \zeta_{l}^{B}(t), \quad l=1, 2,...,N.
$$
Obtaining $\nu_{k}(t)$ from the first equation and substituting it into the rest ones gives
\begin{equation}
\begin{array}{l}
\displaystyle (E-\omega_{0}) \zeta_{l}^{A}(t)=-\int_{0}^{\infty} \frac{J(\omega) d\omega}{\omega-E} \sum_{l=1}^{N} \zeta_{l}^{A}(t)-\int_{0}^{\infty} \frac{J'(\omega) d\omega}{\omega-E} \sum_{l=1}^{N} \zeta_{l}^{B}(t),\\\\
\displaystyle (E-\omega_{0}) \zeta_{l}^{B}(t)=-\int_{0}^{\infty} \frac{J'(\omega) d\omega}{\omega-E} \sum_{l=1}^{N} \zeta_{l}^{A}(t)-\int_{0}^{\infty} \frac{J(\omega) d\omega}{\omega-E} \sum_{l=1}^{N} \zeta_{l}^{B}(t).
\end{array}
\end{equation}
Consequently, by eliminating the amplitudes, these equations can be combined in a compact form as
\begin{eqnarray}
E=\omega_{0}-N(1+\theta) \int_{0}^{\infty} \frac{J(\omega) d\omega}{\omega-E}.
\end{eqnarray}
It is obvious that the solution of Eq. (20) highly depends on the particular choice of the spectral density $J(\omega)$ of the reservoir, the number of atoms $N$ and
the SGI parameter $\theta$. The existence of a bound state in the spectrum of Eq. (17) requires that Eq. (20) must have at least a real solution in the negative energy range, i.e. $E < 0$.
On the other hand, it has only complex solutions when a bound state is not formed and the corresponding eigenstate
experiences decay from the imaginary part of the eigenvalue during the time evolution, so the excited-state population approaches to zero asymptotically.
However, when a bound state is formed the population of the atomic excited state is constant in time because it has a
vanishing decay rate during the time evolution. Therefore, in the next section, we demonstrate the constructive role of
the system-reservoir bound states on the two-qutrit entanglement protection through additional qutrits and the SGI parameter
without manipulating the spectral density of the reservoirs.

\section{Two-qutrit entanglement protection}
In this section, we consider two V-type atoms each of which is embedded in independent reservoirs along with $N-1$ additional atoms, as depicted in Fig. 1(b).
Before obtaining the dynamics of the two-qutrit V-type system, it is very instructive to rewrite the dynamics of the single-qutrit described in Sec. II,
in terms of the well-known Krauss representation.
To this aim, the first qutrit ($l=1$) is considered as our main concern of single-qutrit system and the $N-1$ remainder ones are considered as the additional qutrits.
As a result, by considering $\zeta_{l}^{A}(0)=\zeta_{l}^{B}(0)=0$ with $l\neq 1$, the probability amplitudes $\zeta_{1}^{A}(t)$ and $\zeta_{1}^{B}(t)$
take the following simple form
\begin{equation}
\begin{array}{l}
\zeta_{1}^{A}(t)=G_{1}(t) \zeta_{1}^{A}(0)+G_{2}(t) \zeta_{1}^{B}(0),\\\\
\zeta_{1}^{B}(t)=G_{2}(t) \zeta_{1}^{A}(0)+G_{1}(t) \zeta_{1}^{B}(0),
\end{array}
\end{equation}
where
\begin{equation}
\begin{array}{l}
\displaystyle G_{1}(t)=\frac{\mathcal{G}_{+}(t)+\mathcal{G}_{-}(t)}{2}+\frac{N-1}{2N} \bigg(2-\mathcal{G}_{+}(t)-\mathcal{G}_{-}(t) \bigg),\\\\
\displaystyle G_{2}(t)=\frac{\mathcal{G}_{+}(t)-\mathcal{G}_{-}(t)}{2}+\frac{N-1}{2N} \bigg(-\mathcal{G}_{+}(t)+\mathcal{G}_{-}(t) \bigg).
\end{array}
\label{eq:xdef}
\end{equation}
It should be noted that, in the limit $N\rightarrow \infty$, we have $G_{1}\rightarrow 1$ and $G_{2}\rightarrow 0$ and therefore $\zeta_{1}^{A}(t)\rightarrow \zeta_{1}^{A}(0)$
and $\zeta_{1}^{B}(t)\rightarrow \zeta_{1}^{B}(0)$, which means that the initial condition is preserved as the number of qutrits becomes large.
Using Eq. (6), the explicit form of the reduced density operator of the $1^{th}$ atom in the basis $\{|A\rangle, |B\rangle, |C\rangle\}$,
can be obtained by tracing over the reservoir and the other atoms as follows
\begin{eqnarray}
\varrho_{1}(t)=\left(
                \begin{array}{ccc}
                  |\zeta_{1}^{A}(t)|^2 & \zeta_{1}^{A}(t) \zeta_{1}^{B}(t)^{*} & \zeta_{1}^{A}(t) \zeta_{0}^{*}\\\\
                  \zeta_{1}^{B}(t) \zeta_{1}^{A}(t)^{*} & |\zeta_{1}^{B}(t)|^2 & \zeta_{1}^{B}(t) \zeta_{0}^{*}\\\\
                  \zeta_{0} \zeta_{1}^{A}(t)^{*} & \zeta_{0} \zeta_{1}^{B}(t)^{*} & 1-|\zeta_{1}^{A}(t)|^2-|\zeta_{1}^{B}(t)|^2\\
                \end{array}
              \right).
\end{eqnarray}
Therefore, $\varrho_{1}(t)$ can be written in terms of the Kraus representation as
\begin{eqnarray}
\varrho_{1}(t)=\sum_{i=1}^{3}\mathcal{K}_{i}\varrho_{1}(0)\mathcal{K}^{\dagger}_{i},
\end{eqnarray}
with $\sum_{i=1}^{3}\mathcal{K}^{\dagger}_{i}\mathcal{K}_{i}=I_{3}$, where $I_{3}$ is identity operator on the Hilbert space of a three-level system and
$$
\mathcal{K}_{1}=\left(
                    \begin{array}{ccc}
                      G_{1}(t) & G_{2}(t) & 0 \\
                      G_{2}(t) & G_{1}(t) & 0 \\
                      0 & 0 & 1 \\
                    \end{array}
                  \right),
$$
\begin {equation}\label{}
\mathcal{K}_{2}=\sqrt{\frac{1-|G_{1}(t)-G_{2}(t)|^{2}}{2}}\left(
                    \begin{array}{ccc}
                      0 & 0 & 0 \\
                      0 & 0 & 0 \\
                      1 & -1 & 0 \\
                    \end{array}
                  \right),
\end {equation}
$$
\mathcal{K}_{3}=\sqrt{\frac{1-|G_{1}(t)+G_{2}(t)|^{2}}{2}}\left(
                    \begin{array}{ccc}
                      0 & 0 & 0 \\
                      0 & 0 & 0 \\
                      1 & 1 & 0 \\
                    \end{array}
                  \right),
$$
are the related Kraus operators.
Now, we can easily extend the above method for dynamics of a system consists of two identical V-type atoms each of which independently interact with a reservoir. Let's $\rho_{S}(0)$ be defined as a density matrix of a two-qutrit system on the $3\otimes3$ Hilbert space. So in this way, its time development at time $t$ becomes
\begin{eqnarray}
\rho_{S}(t)=\sum_{k,l=1}^{3}\mathcal{K}_{k,l}\rho_{S}(0)\mathcal{K}^{\dagger}_{k,l}, \quad \sum_{k,l=1}^{3}\mathcal{K}^{\dagger}_{k,l}\mathcal{K}_{k,l}=I_{3}\otimes I_{3},
\end{eqnarray}
where $\mathcal{K}_{k,l}=\mathcal{K}_{k}\otimes \mathcal{K}_{l}$.
We suppose that the two-qutrit system is initially prepared in a maximally entangled state as $|\psi_{0}\rangle=\frac{1}{\sqrt{3}}(|00\rangle+|11\rangle+|22\rangle)$.
Therefore, the time evolved two-qutrit system can be obtained by using Eq. (26). In what follows, we use the negativity as a suitable
measure to quantify the amount of entanglement of the two-qutrit V-type system defined as \cite{Vidal}
\begin{eqnarray}
\mathcal{N}(\rho)=\frac{\|\rho^{T_{S_{1}}}\|-1}{2},
\end{eqnarray}
where $\rho^{T_{S_{1}}}$ is the partial transpose with respect to the subsystem 1 and $\|. \|$ denotes the trace norm.
Equivalently, the negativity is the absolute value of sum of the negative eigenvalues of the partially transposed density matrix.

The time evolution of the negativity for the two-qutrit V-type system where the two distant qutrits are evolving on their own reservoirs has been shown in Figs. 2(a) and 2(b).
Surprisingly, it is observed that without the presence of the additional qutrits (that is, the case when $N=1$) and by considering the fixed condition of the reservoirs
($\gamma_{0}=1$ (in units of $\omega_{0}$) and $\lambda=0.8$ (in units of $\omega_{0}$)), we are faced with two different behaviors:
the negativity eventually decays to zero with a weak SGI ($\theta=0.5$) where the bound state in the system-environment spectrum is not formed, as shown by the solid line in Fig. 2 (c),
but for the strongest SGI ($\theta=1$), it approaches asymptotically to a non-zero steady value where the formation of bound state in the system-environment spectrum is taken place as shown by the solid line in Fig. 2 (d). Also, as we can see from Figs. 2(c) and 2(d), the bound states are formed with higher degree of boundedness by inserting more additional qutrits into the reservoirs which in turns, provides better protection of entanglement from sudden death.
Moreover, it should be noted that although by entering the additional qutrits into the reservoir, the system-reservoir bound states become stronger in general, however, they are more considerable for the strongest SGI ($\theta=1$). By this reason, the two-qutrit entanglement is well-protected as the number of additional qutrits grows for $\theta=1$ (strong SGI) in comparison to the case $\theta=0.5$ (weak SGI).

For better viewing of the dependencies of the negativity on the two parameters $\gamma_{0}$ and $\theta$, we plot Figs. 3(a), 3(b), 3(c) and 3(d) to show the protection process of the entanglement in terms of the parameters $\gamma_{0}$ (in units of $\omega_{0}$) and $\theta$ in the absence and presence of additional qutrits.
It is obvious that the maximum amount of entanglement is occurred when $\gamma_{0}=0$, which is relevant to the situation that the system does not interact
with their environment and therefore the initial entanglement does not decay.
All of these observations returns to the fact that the existence of system-reservoir bound state with higher degree of boundedness ensures the better protection of entanglement from dissipative noises.

\section{Conclusion}
In summary, we investigated the protection of two-qutrit entanglement between two V-type atoms each of which is contained independently in a dissipative reservoir. It was demonstrated that the protection process is related decisively into the formation of the system-reservoir bound states. In other words, existence of stronger system-reservoir bound state leads to the better protection of two-qutrit entanglement. It was shown that the degree of boundedness of the system-reservoir bound states is well-controlled by entering the additional qutrits into the reservoirs, and also through the manipulation of SGI of decay channels of the V-type atoms.

\newpage
Fig. 1. (a) A V-type three-level atom immersed in a dissipative reservoir in the presence of 5 ($N-1$) similar additional atoms. (b) Two initially entangled V-type atoms, each of which is contained in a dissipative reservoir in the presence of 5 ($N-1$) similar additional atoms.
\begin{figure}
\centering
    \centering
        \qquad \qquad\qquad \qquad (a) \qquad\qquad \quad\qquad\qquad\qquad\\{
        \includegraphics[width=4.5in]{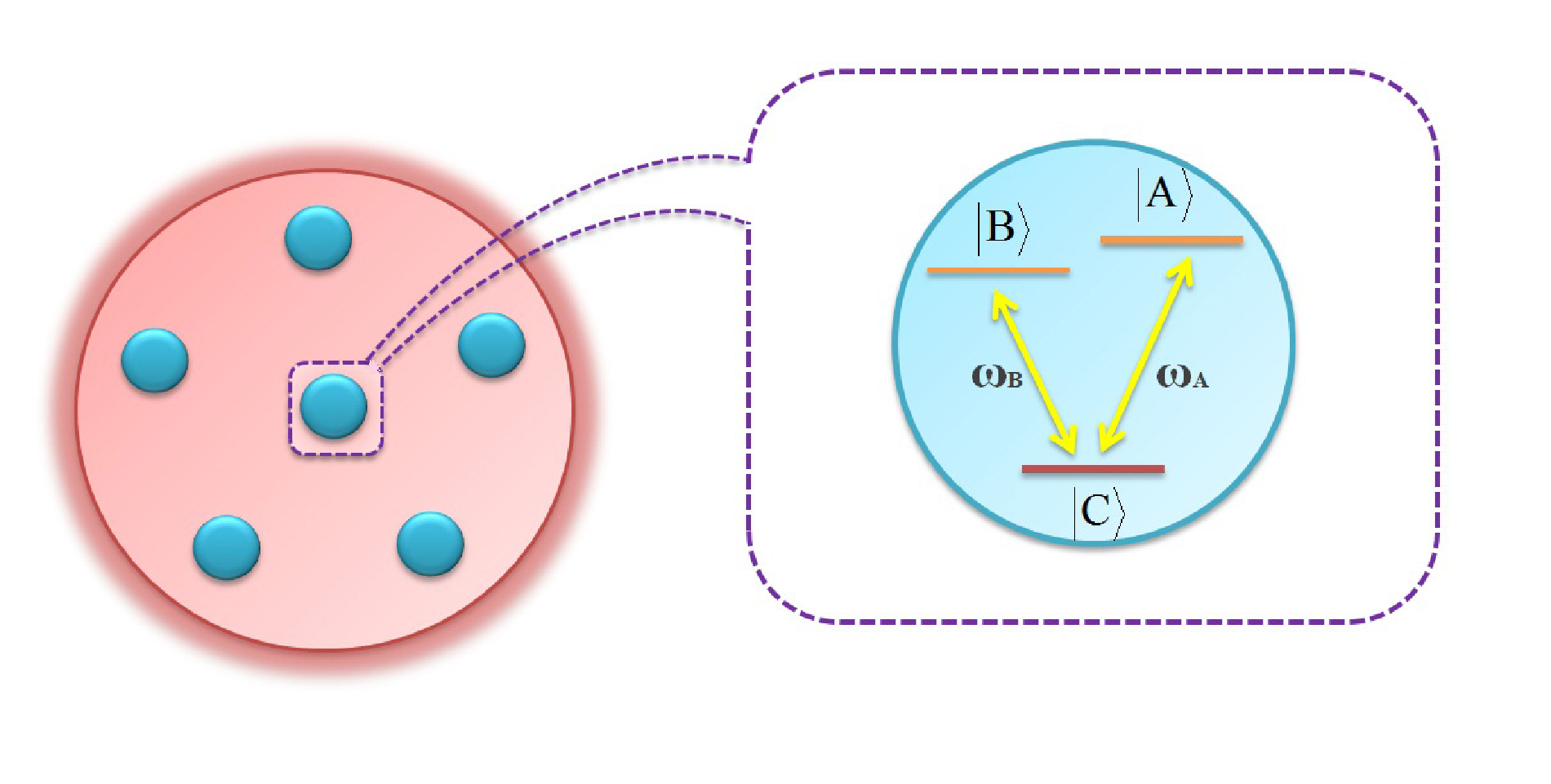}
        \label{fig:first_sub}
    }
    \\
    \centering
       \qquad \qquad\qquad \qquad (b) \qquad\qquad \quad\qquad\qquad\qquad\\{
        \includegraphics[width=4.5in]{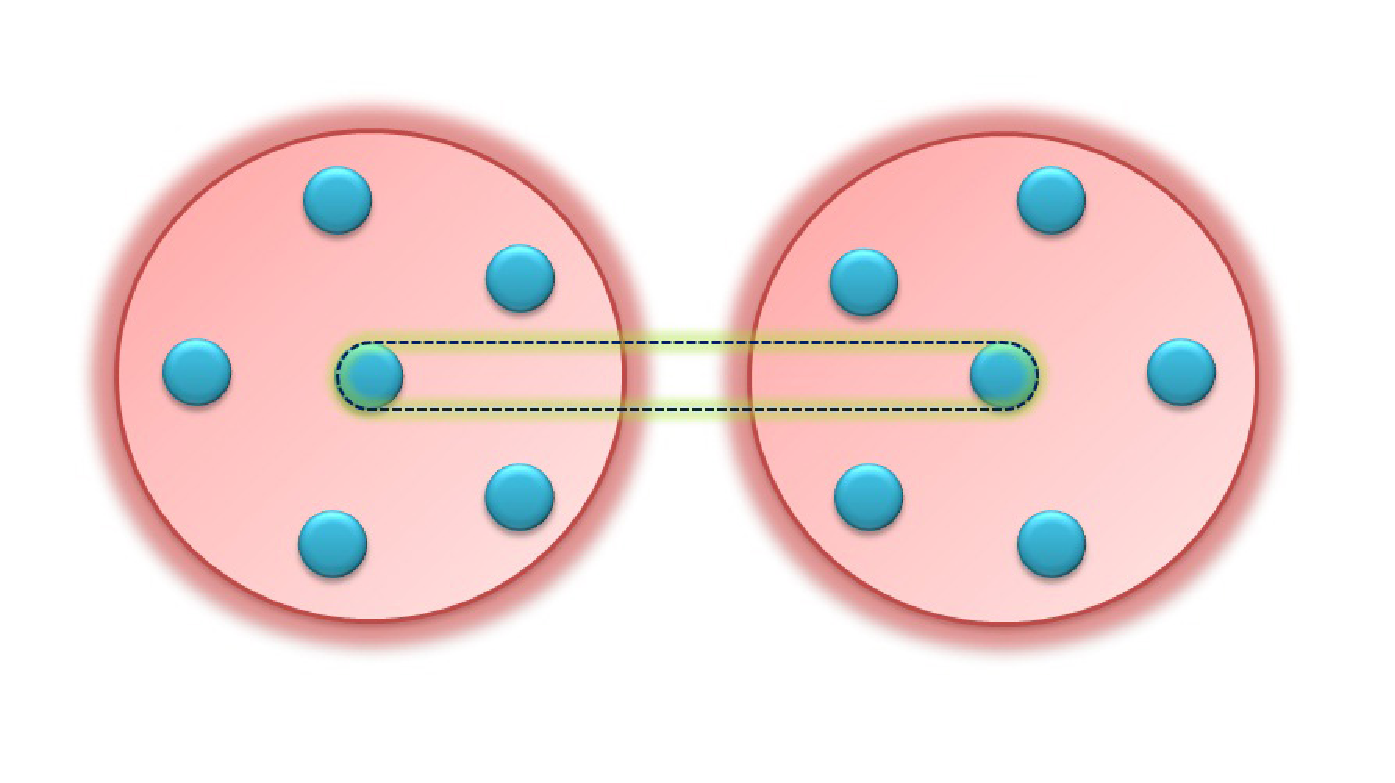}
        \label{fig:second_sub}
    }\\
    \centering
    \caption{}
\end{figure}

\newpage
Fig. 2. (a, b) The negativity for $\gamma_{0}=1$ (in units of $\omega_{0}$) and $\lambda=0.8$ (in units of $\omega_{0}$) versus time (in units of $\omega_{0}^{-1}$).
(c, d) The negative energy spectrum of the system-reservoir, E (in units of $\omega_{0}$), in terms of the coupling
strength $\gamma_{0}$ (in units of $\omega_{0}$), for the reservoir with Lorentizan spectral density.
The panels (a, c) are plotted with $\theta=0.5$ and (b, d) with $\theta=1$.

\begin{figure}
        \qquad \qquad\qquad\qquad \qquad a \qquad\qquad \quad\qquad\qquad\qquad\qquad\qquad\qquad b\\{
        \includegraphics[width=3in]{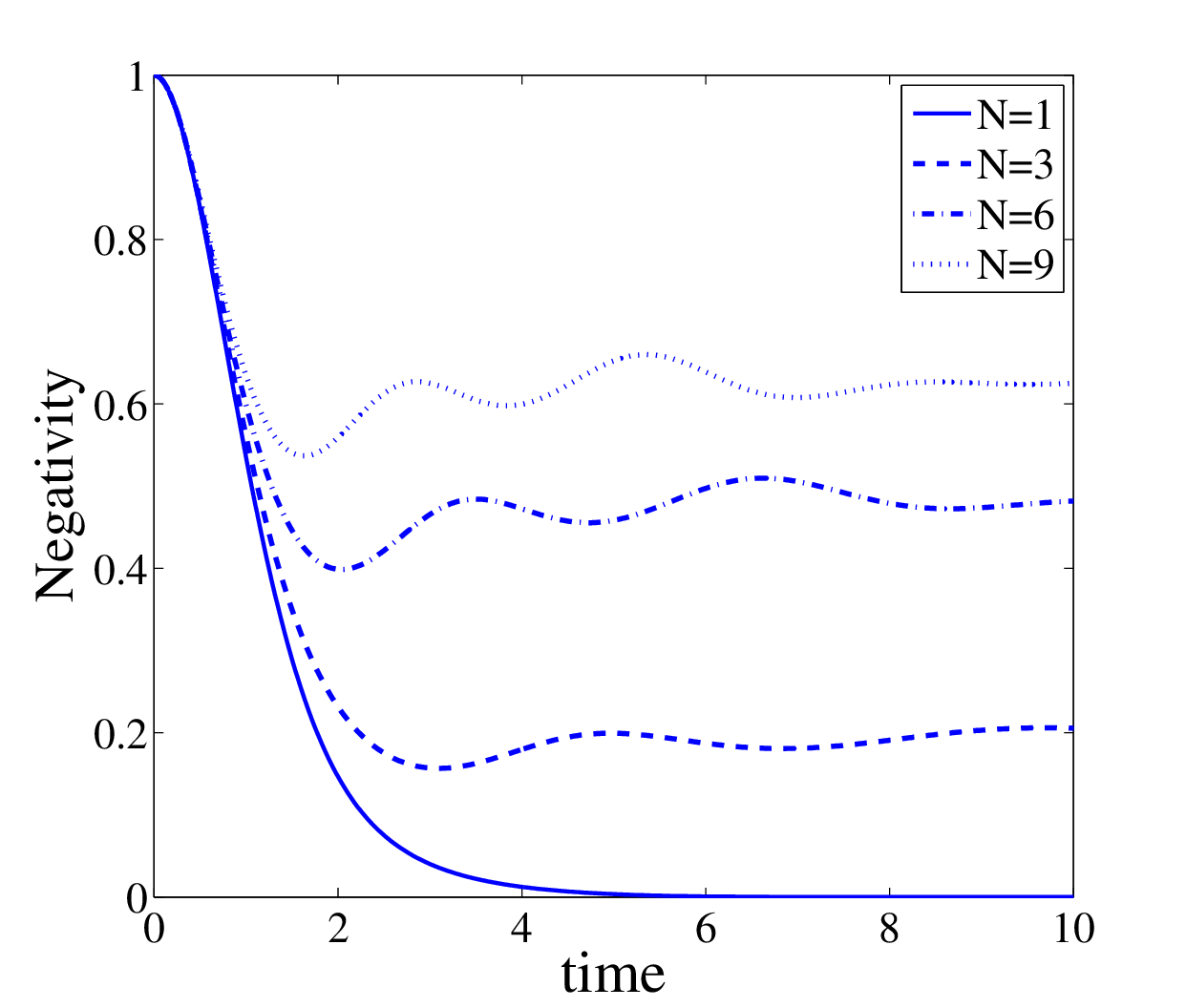}
        \label{fig:first_sub}
    }{
        \includegraphics[width=3in]{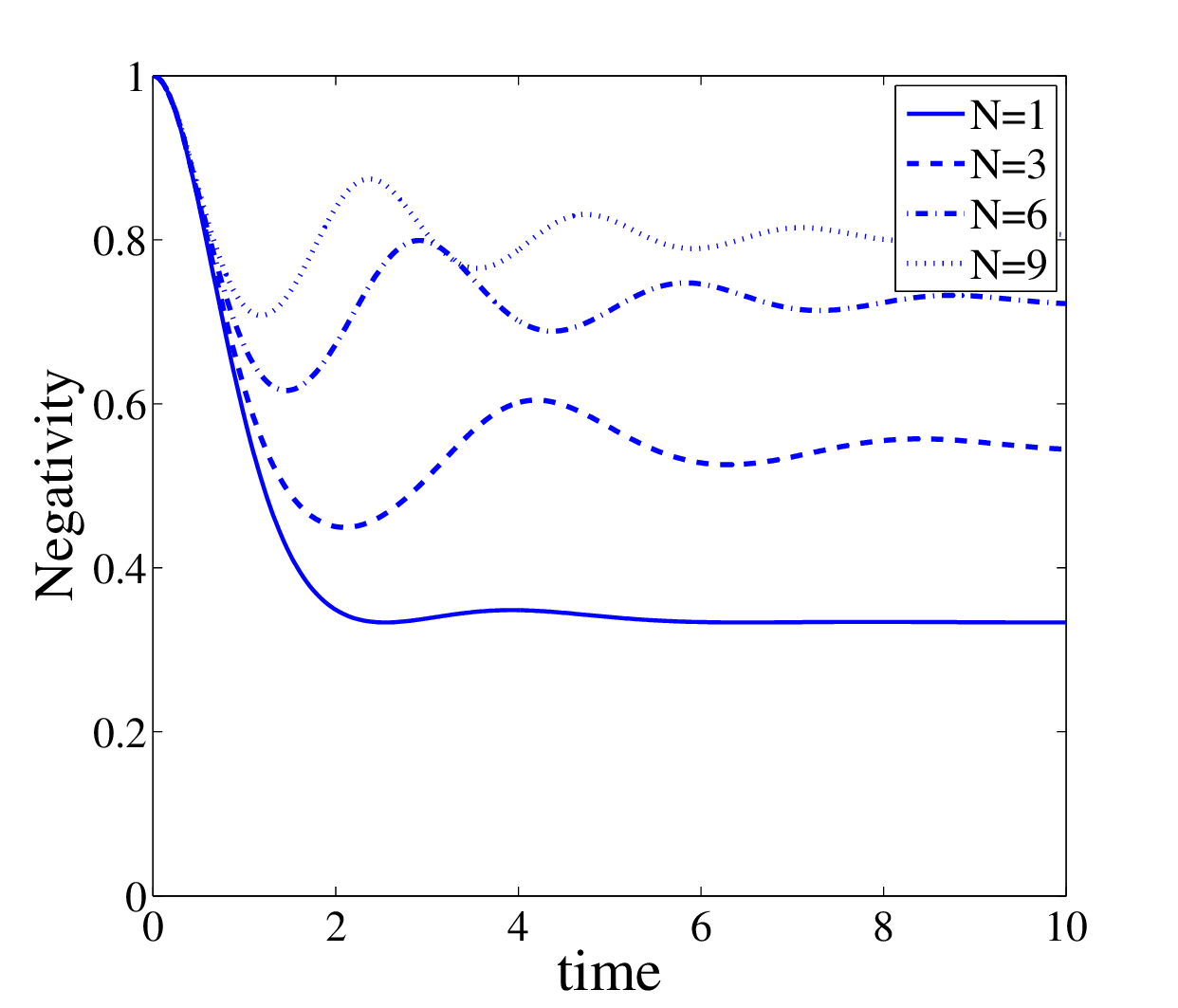}
        \label{fig:second_sub}
    }\\ \par \quad \quad\qquad\qquad\qquad \qquad c \qquad \qquad\qquad\qquad\qquad\quad\qquad\qquad\qquad d\\{
        \includegraphics[width=3in]{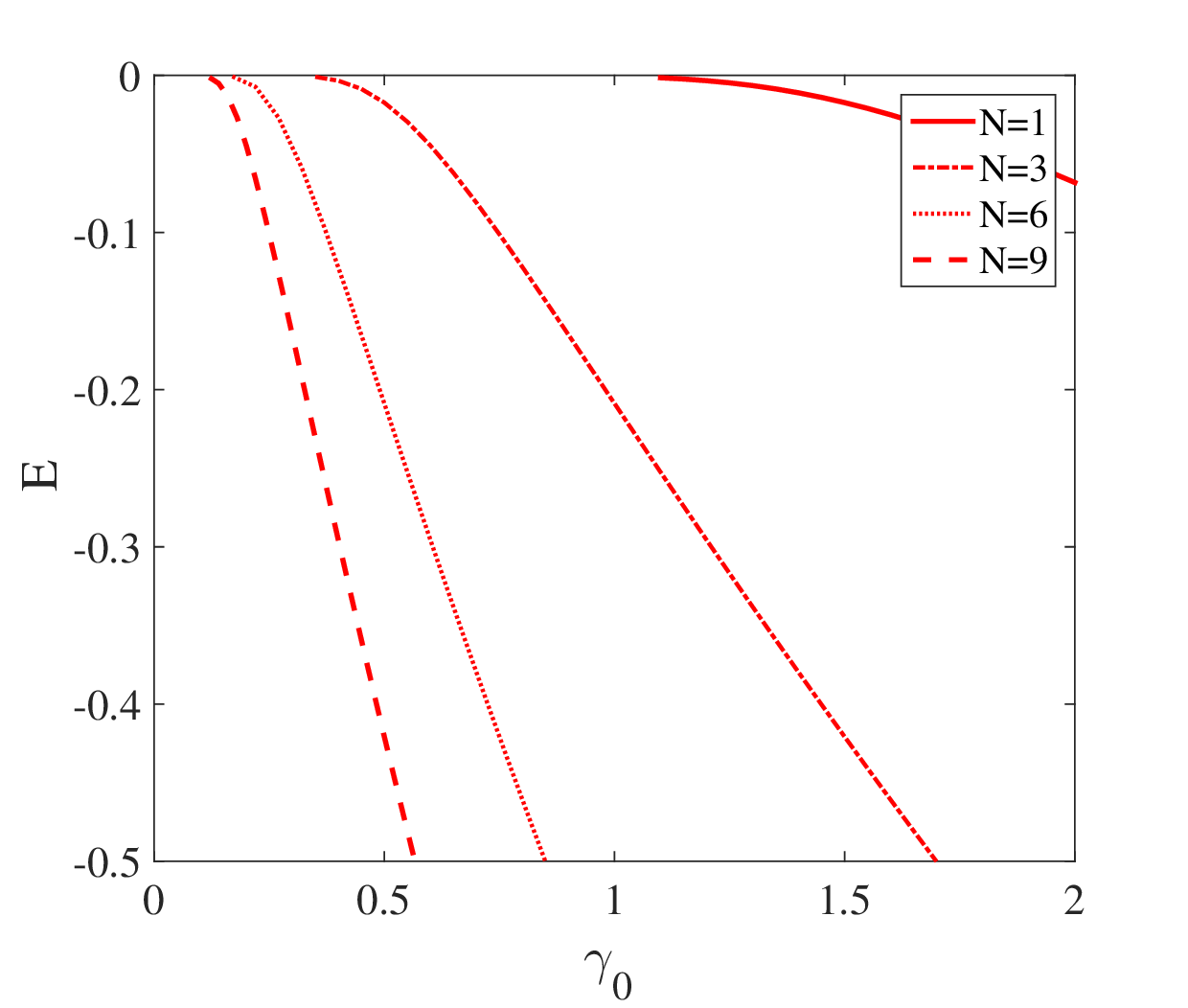}
        \label{fig:first_sub}
    }{
        \includegraphics[width=3in]{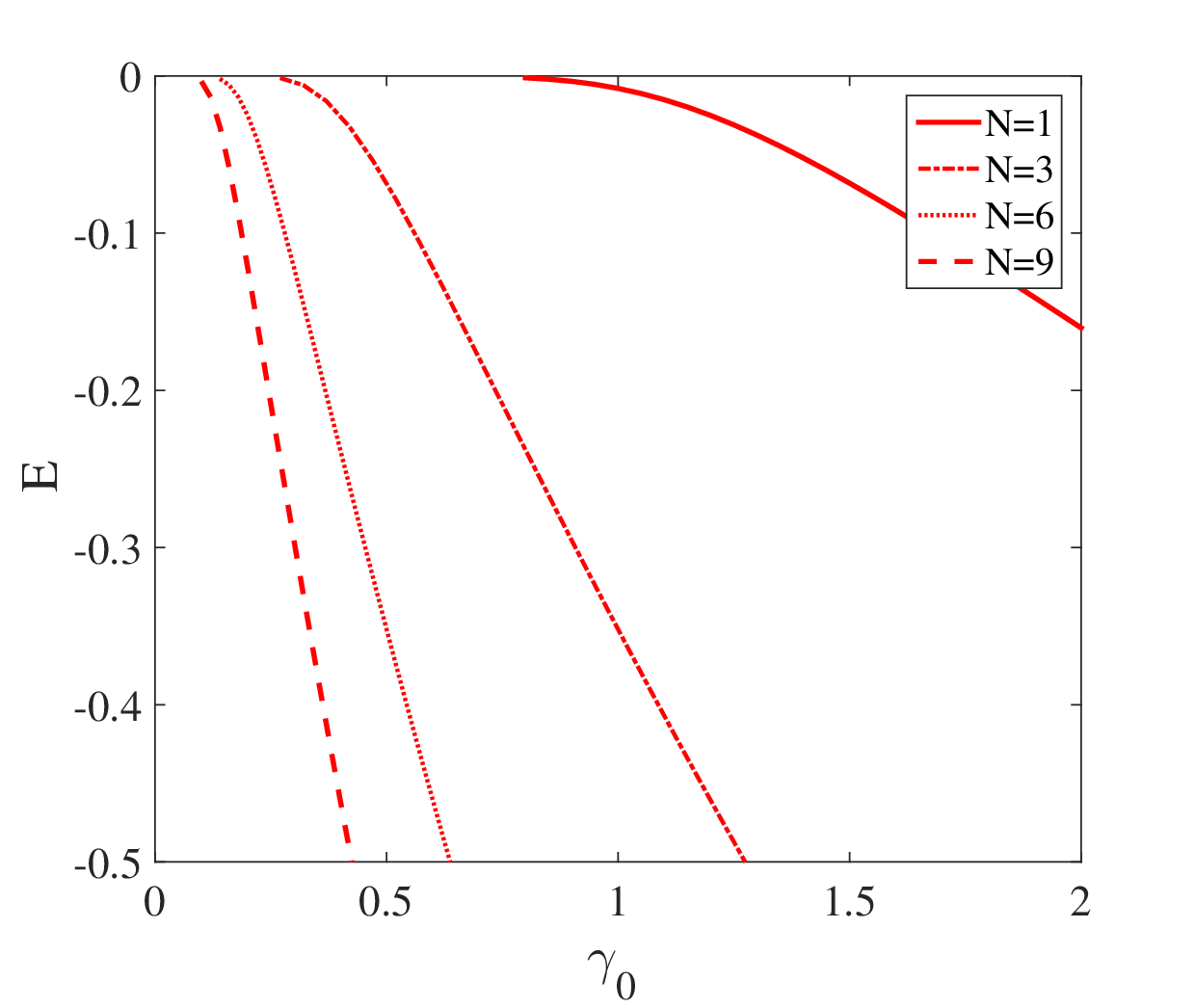}
        \label{fig:second_sub}
    }
    \caption{}
    \end{figure}

\newpage
Fig. 3. The negativity in terms of the $\gamma_{0}$ (in units of $\omega_{0}$) and $\theta$ with (a) $N=1$, (b) $N=3$, (c) $N=6$ and (d) $N=9$, by considering that $\lambda=0.8$ (in units of $\omega_{0}$) and $t=10$ (in units of $\omega_{0}^{-1}$).

\begin{figure}
        \qquad \qquad\qquad\qquad \qquad a \qquad\qquad \quad\qquad\qquad\qquad\qquad\qquad\qquad b\\{
        \includegraphics[width=3in]{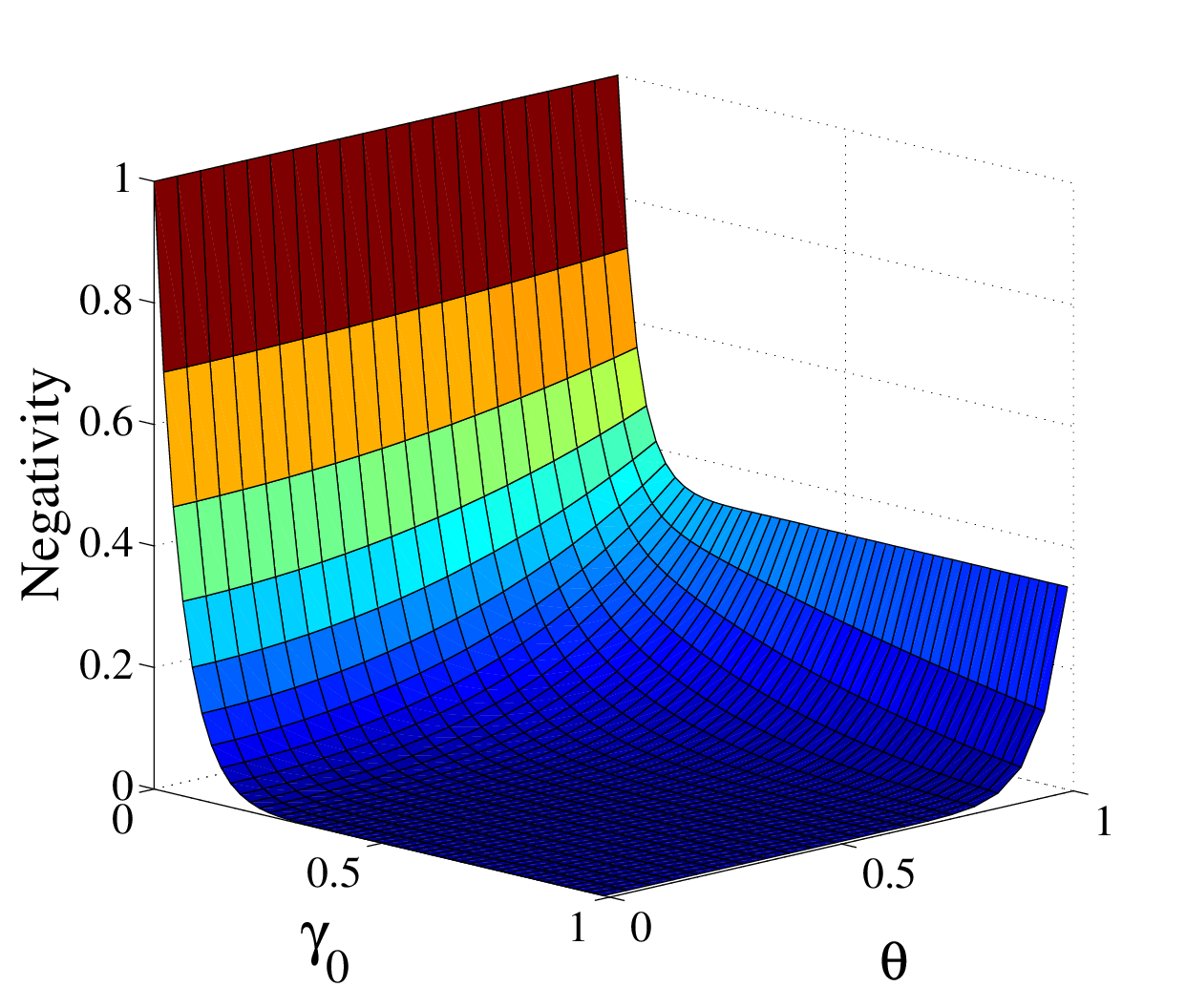}
        \label{fig:first_sub}
    }{
        \includegraphics[width=3in]{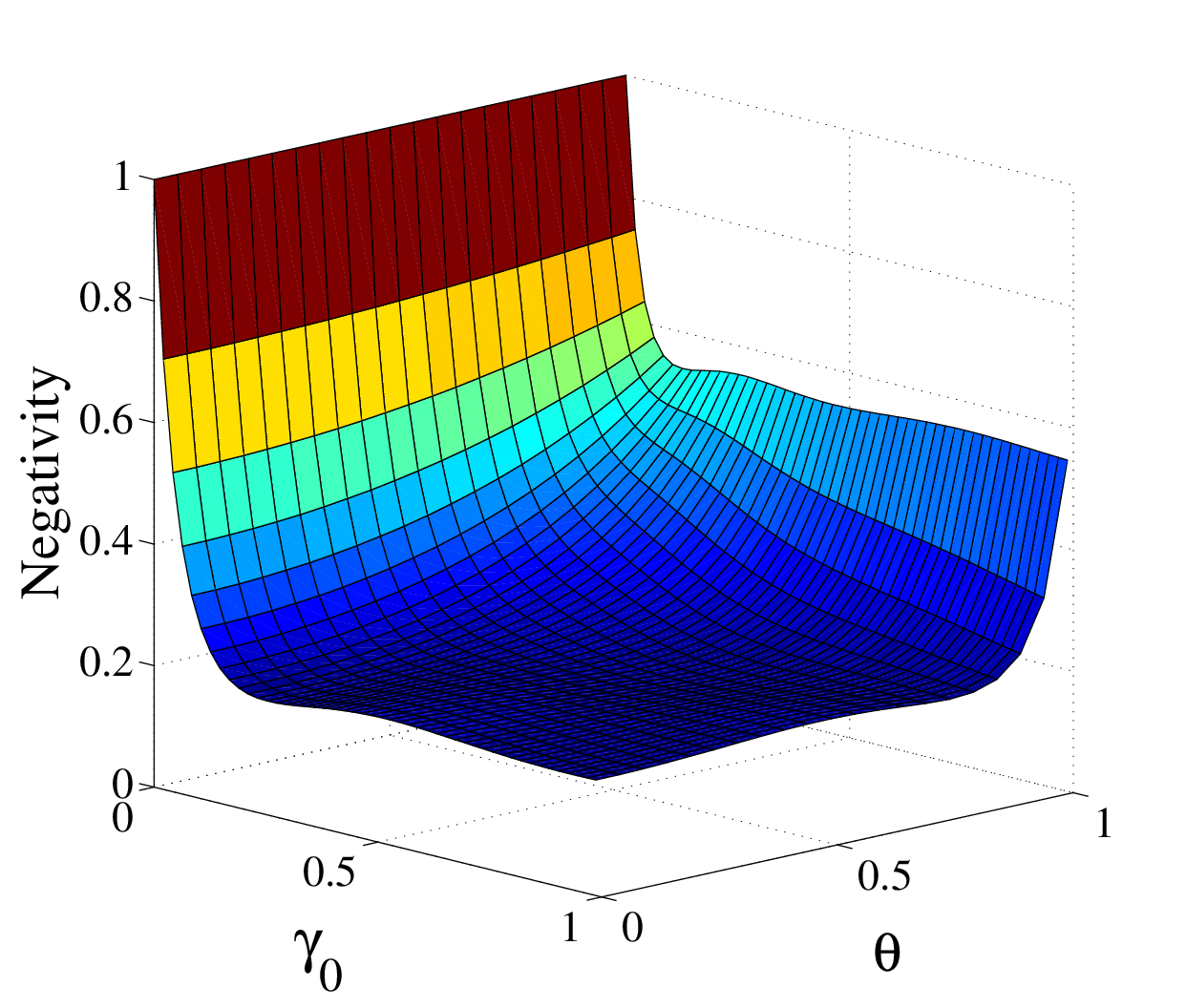}
        \label{fig:second_sub}
    }\\ \par \quad \quad\qquad\qquad\qquad \qquad c \qquad \qquad\qquad\qquad\qquad\quad\qquad\qquad\qquad d\\{
        \includegraphics[width=3in]{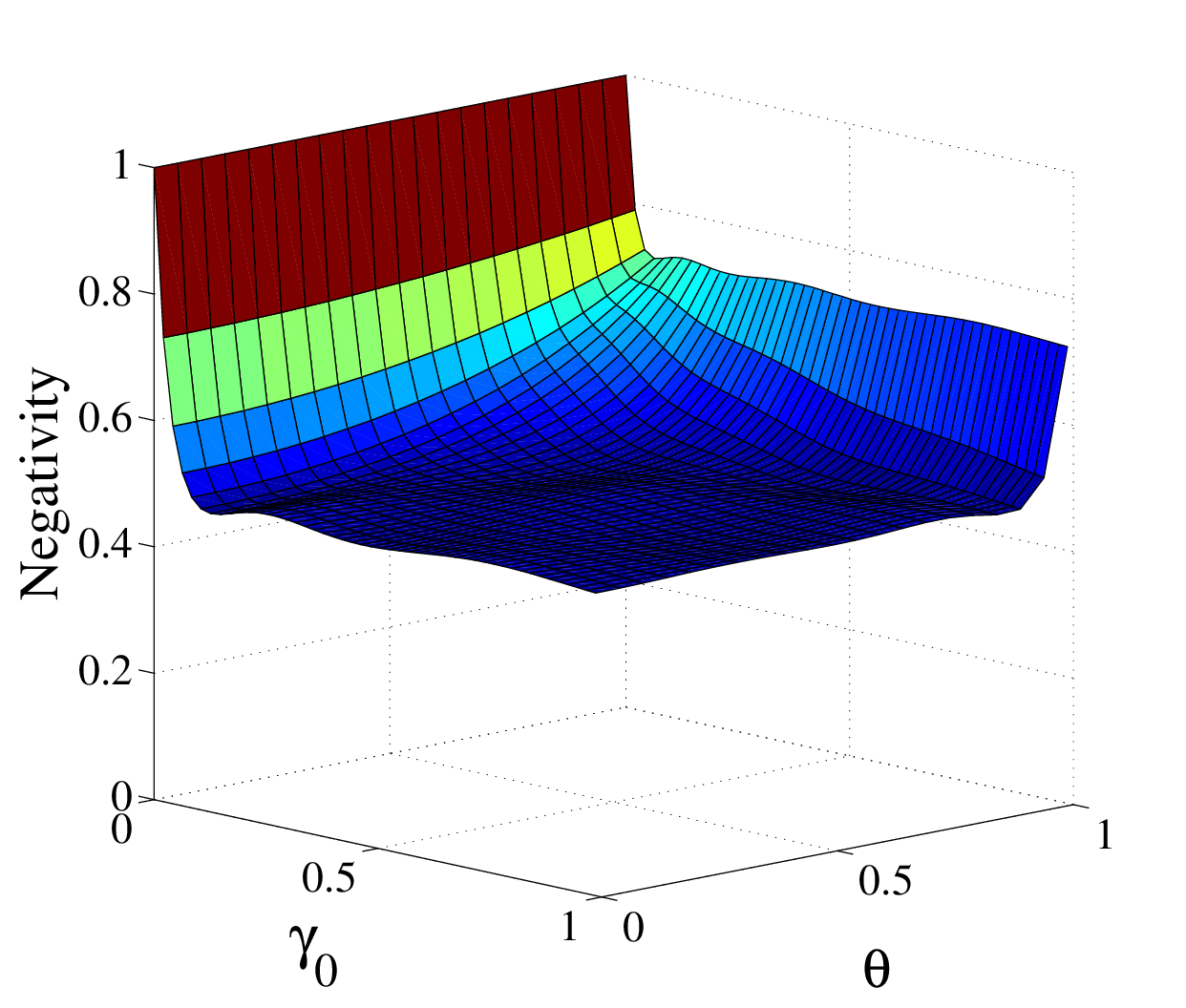}
        \label{fig:first_sub}
    }{
        \includegraphics[width=3in]{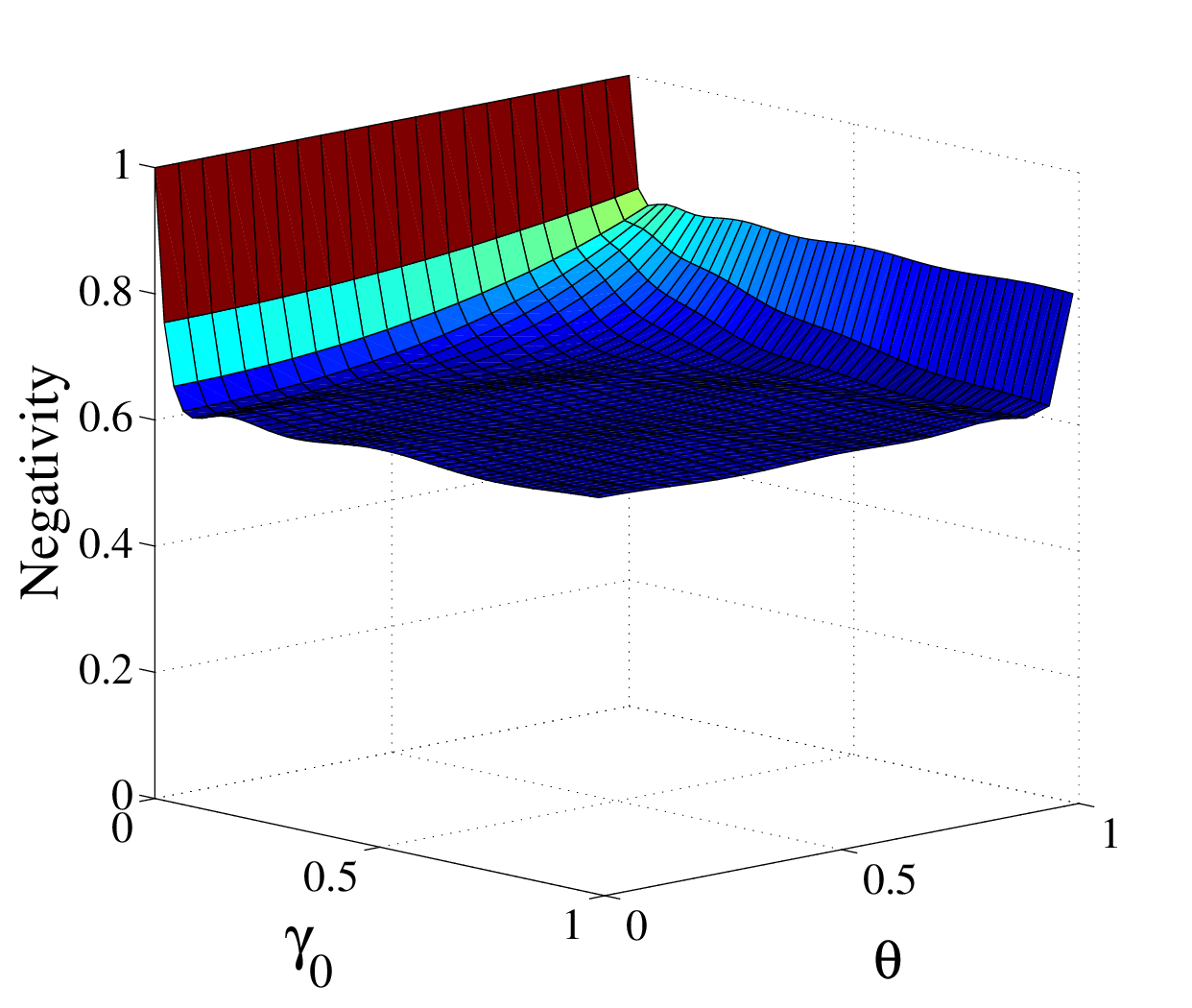}
        \label{fig:second_sub}
    }
    \caption{}
    \end{figure}


\newpage
\begin{thebibliography}{99}

\bibitem{Horodecki}
R. Horodecki, P. Horodecki, M. Horodecki and K. Horodecki, Rev. Mod. Phys. 81, 865 (2007).

\bibitem{Maniscalco}
S. Maniscalco, F. Francisca, R. Zaffino, N. Gullo and F. Plastina, Phys. Rev. Lett. 100, 090503 (2008).

\bibitem{Mundarain}
D. Mundarain and M. Orszag, Phys. Rev. A. 79, 052333 (2009).

\bibitem{Rossi}
R. Rossi Jr, Phys. Lett. A. 374, 2331 (2010).

\bibitem{Hou}
Y. Hou, G. Zhang, Y. Chen and H. Fan, Ann. Phys. 327, 292 (2012).

\bibitem{Bellomo}
B. Bellomo, R. L. Franco, S. Maniscalco and G. Compagno, Phys. Rev. A. 78, 060302 (2008).

\bibitem{Xiao}
X. Xiao, Y. Li, K. Zeng and C. Wu, J. Phys. B: At. Mol. Opt. Phys. 42, 235502 (2009).

\bibitem{An2}
N. Ba An, J. Kim and K. Kim, Phys. Rev. A. 84, 022329 (2011).

\bibitem{An3}
N. Ba An, Phys. Lett. A. 337, 2520 (2013).

\bibitem{Bourennane}
M. Bourennane, A. Karlsson and G. Bjork, Phys. Rev. A. 64, 012306 (2001).

\bibitem{Kaszlikowski}
D. Kaszlikowski, P. Gnacinski, M. Zukowski, W. Miklaszewski and A. Zeilinger, Phys. Rev. Lett. 85, 4418 (2000).

\bibitem{Harris}
S. E. Harris, Phys. Rev. Lett. 62, 1033 (1989).

\bibitem{Scully}
M. O. Scully, S. Y. Zhu and A. Gavrielides, Phys. Rev. Lett. 62, 2813 (1989).

\bibitem{Lambropoulos}
P. Lambropoulos, G. M. Nikolopoulos, T. R. Nielsen and S. Bay, Rep. Prog. Phys. 63, 455 (2000).

\bibitem{Tong}
Q. Tong, J. An, H. Luo and C. Oh, Phys. Rev. A. 81, 052330 (2010).

\bibitem{Vidal}
G. Vidal, R. F. Werner, Phys. Rev. A. 65, 032314 (2002).


\end{thebibliography}
\end{document}